\documentclass{article}
\usepackage{spconf,graphicx,hyperref}
\usepackage{amsmath,amssymb,amsfonts}
\usepackage{makecell, multirow}
\usepackage[compress]{cite}
\usepackage{tikz}
\usetikzlibrary{shapes,arrows,positioning}
\usepackage{color}

\renewenvironment{thebibliography}[1]{%
   \begin{oldthebibliography}{#1}%
     \setlength{\itemsep}{2pt}%
}%
{%
   \end{oldthebibliography}%
}
\title{Beyond Omnidirectional: Neural Ambisonics Encoding for Arbitrary Microphone Directivity Patterns using Cross-Attention\vspace{-2.5pt}}
\name{Mikko Heikkinen$^{\dag}$ \qquad Archontis Politis$^{\star}$ \qquad Konstantinos Drossos$^{\ddag}$ \qquad Tuomas Virtanen$^{\star}$\vspace{-3pt}}
\address{$^{\dag}$$^{\ddag}$ Nokia, Tampere$^{\dag}$ / Espoo$^{\ddag}$, Finland \\
    $^{\star}$ Tampere University, Tampere, Finland\vspace{-3pt}}
\begin{document}
%
\maketitle
\begin{abstract}
We present a deep neural network approach for encoding microphone array signals into Ambisonics that generalizes to arbitrary microphone array configurations with fixed microphone count but varying locations and frequency-dependent directional characteristics. Unlike previous methods that rely only on array geometry as metadata, our approach uses directional array transfer functions, enabling accurate characterization of real-world arrays. The proposed architecture employs separate encoders for audio and directional responses, combining them through cross-attention mechanisms to generate array-independent spatial audio representations. We evaluate the method on simulated data in two settings: a mobile phone with complex body scattering, and a free-field condition, both with varying numbers of sound sources in reverberant environments. Evaluations demonstrate that our approach outperforms both conventional digital signal processing-based methods and existing deep neural network solutions. Furthermore, using array transfer functions instead of geometry as metadata input improves accuracy on realistic arrays.
\end{abstract}
\begin{keywords}
Spatial Audio, Ambisonics, Microphone Array
\end{keywords}
\vspace{-7pt}
\section{Introduction}
\label{sec:intro}\vspace{-5pt}
Ambisonics is a device-independent spatial audio format widely adopted in immersive communication, virtual/extended reality, and standardized in 3GPP IVAS, MPEG-I, and AOM immersive audio formats~\cite{zotter2019ambisonics,olivieri2019scene,pulkki2018parametric, multrus24-AES,herre2024mpeg,hwang2023iamf}. While spherical microphone arrays (SMAs) represent the optimal arrangement for Ambisonics audio capture, consumer devices often employ compact and irregular microphone arrays (MAs) due to design constraints dictated by the device's primary function.

Traditional Ambisonics encoding implements a time-invariant filter matrix, obtained by solving the optimal matching problem between the MA's array transfer functions (ATFs) and spherical harmonics (SH) in a least squares sense. This approach, while well-established for SMAs~\cite{zotter2019ambisonics, jin2013design, zotkin2017incident, politis2017comparing}, can be extended to irregular MAs~\cite{laborie2003new, zotkin2017incident, jin2013design, politis2017comparing, bastine2022ambisonics}. These model-based methods, which consider MA geometry, can readily provide encoding solutions for different array configurations. However, they are fundamentally limited by the physical constraints of using few microphones to capture Ambisonic components, typically exhibiting energy loss or noise amplification in first- or higher-order channels at low frequencies, and distorted SH patterns above the spatial aliasing frequency. These limitations are directly related to the MA's size and configuration~\cite{rafaely2015fundamentals}.

Recent advances in Ambisonics encoding include signal-dependent parametric methods that adaptively transform MA signals using estimated spatial parameters~\cite{mccormack2022parametric}. While more flexible geometrically, their performance depends on how well the sound scene matches their assumed parametric model. Non-linear signal-dependent approaches using deep neural networks (DNNs) have shown promising results, improving encoding performance both in low frequencies and above spatial aliasing~\cite{gao2022sparse,heikkinen2024neural,qiao2025neural}. Some recent works in MA-to-binaural conversion~\cite{hsu2023model,hsu2024tunable}, have shown partial cross-device generalization enabled by geometry-aware features, but as a rule DNN methods are typically limited by requiring retraining for each MA geometry. 
Our recent work eliminates need for retraining by incorporating microphone locations as additional metadata alongside the input signals in its DNN model~\cite{heikkinen2025gen}. This approach enables generalization to unseen MA configurations during inference. However, it is limited to a free-field setting of omnidirectional arrays in its modeling capability because the location metadata cannot convey more detailed information about the ATFs.

This paper presents a DNN model that processes MA signals and complex-valued directional ATFs to predict Ambisonic output, with generalization to unseen arrays; specifically, microphone arrays with the same number of microphones but arbitrary ATFs that encode microphone location, directivity patterns, and device body scattering effects. The architecture features separate signal and directivity metadata encoders connected through cross-attention, producing an array-independent latent representation that is decoded into a filter matrix for MA-to-Ambisonics transformation.
The model is evaluated using simulated data in a mobile-phone setting with complex device body scattering and in a free-field environment. Comparisons against traditional static encoding, a parametric method, and an existing DNN approach demonstrate that in complex scattering scenarios the proposed method achieves the highest scale-invariant signal-to-distortion-ratio (SI-SDR) and consistently outperforms the static encoder across all Ambisonics signal metrics. In free-field conditions, the model achieves the best SI-SDR and performs comparably to existing neural solutions, while also surpassing the static encoder. Analysis of attention weights provides insights into the model's utilization of directivity metadata for spatial encoding.

\section{Proposed Method}
\label{sec:method}
\subsection{Signal model and encoding task}
Considering an MA recording denoted as a tensor of $P$ spectrograms of MA signals $\mathbf{X}\in\mathbb{C}^{P\times F\times T}$ of $F$ frequency bins and $T$ time frames and a tensor of $(N+1)^2$ spectrograms of Ambisonics signals $\mathbf{B}\in\mathbb{C}^{(N+1)^2\times F\times T}$ up to Ambisonics order N that we intend to encode from the microphone array. Both sets of signals are generated through the same sound field characterized by the complex plane wave amplitude distribution $a_{t,f}(\theta,\phi)$ for time index $t$, frequency $f$, azimuth $\theta$, and elevation $\phi$. The acoustical model of these signals is given by ~\cite{politis2017comparing}:
\begin{align}
\mathbf{x}_{t,f}&=\iint a_{t,f}(\theta,\phi) \mathbf{h}_f(\theta,\phi)\sin(\phi)d{\theta}d{\phi} \\
\mathbf{b}_{t,f}&=\iint a_{t,f}(\theta,\phi)\mathbf{y}_N(\theta,\phi)\sin(\phi)d{\theta}d{\phi}
\end{align}
where $\mathbf{x}_{t,f}$ and $\mathbf{b}_{t,f}$ are slices across channels of $\mathbf{X}$ and $\mathbf{B}$ for a time-frequency point $(t,f)$. The ATFs $\mathbf{h}_f \in\mathbb{C}^{P\times 1}$ characterize the directional responses of the array
while the SHs $\mathbf{y}_N \in\mathbb{R}^{(N+1)^2\times 1}$ define Ambisonic directivities. Both $\mathbf{b}$ and $\mathbf{y}_N$ are indexed by SH of order $n$ and degree $m$ up to maximum captured order $N$.

Motivated by the above signal models, we define our task to find a model $\mathcal{M}$ 
\begin{equation}
\mathcal{M}: (\mathbf{X}, \mathbf{H}) \mapsto \mathbf{E}_{t,f}\text{ s.t. } \mathbf{b}_{t,f} \approx \mathbf{E}_{t,f}\mathbf{x}_{t,f} 
\label{eq:task}
\end{equation}
\noindent
that takes as input $\mathbf{X}$ and  tensor of ATFs $\mathbf{H} \in \mathbb{C}^{P \times D \times F_H}$ measured at $D$ discrete directions $(\theta_d,\phi_d)$ and $F_H$ frequency bins, and calculates a mixing matrix $\mathbf{E}\in\mathbb{C}^{(N+1)^2)\times P}$ at each time-frequency point. When $\mathbf{E}$ is applied onto $\mathbf{x}$ it produces a desired approximation $\mathbf{\hat{b}}=\mathbf{E}\mathbf{x}$ to the ground truth.

\subsection{Method}
\label{sec:DNN}

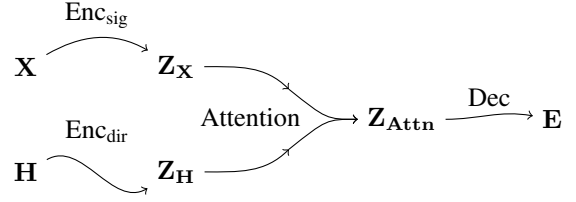
\begin{figure}[!t]
    \centering
    \begin{tikzpicture}
        \node (X) at (0,1.8) {$\mathbf{X}$};
        \node (H) at (0,0.4) {$\mathbf{H}$};
        \node (Zx) at (2,1.8) {$\mathbf{Z_X}$};
        \node (Zh) at (2,0.4) {$\mathbf{Z_H}$};
        \node (Zattn) at (5,1.1) {$\mathbf{Z_{Attn}}$};
        \node (E) at (7,1.1) {$\mathbf{E}$};
        %
        \draw[->] (X) to[out=30,in=150] node[above] {$\text{Enc}_{\text{sig}}$} (Zx);
        \draw[->] (H) to[out=30,in=210] node[above=0.3cm] {$\text{Enc}_{\text{dir}}$} (Zh);
        %
        \draw[->] (Zx) to[out=0,in=135] node[below=0.4cm] {$\text{Attention}$} (3.5,1.5);
        \draw[->] (Zh) to[out=0,in=225] (3.5,0.7);
        \draw[->] (3.5,1.5) .. controls (4,1.1) .. (Zattn);
        \draw[->] (3.5,0.7) .. controls (4,1.1) .. (Zattn);
        %
        \draw[->] (Zattn) to[out=0,in=170] node[above] {$\text{Dec}$} (E);
    \end{tikzpicture}
        \caption{Sketch of the proposed method. The inputs $\mathbf{X}$ and $\mathbf{H}$ are encoded into latent representations $\mathbf{Z_X}$ and $\mathbf{Z_H}$ respectively, which are then combined through an attention mechanism to produce $\mathbf{Z_{Attn}}$. Finally, $\mathbf{Z_{Attn}}$ is decoded to generate the mixing matrix $\mathbf{E}$}
    \label{fig:dnn overview}
    \vspace{-4pt}
\end{figure}

Our method employs a DNN as the model of the task of Eq.~(\ref{eq:task}), which is reformulated as
\begin{align}
    \mathbf{E} &= \text{DNN}(\mathbf{X}, \mathbf{H}),\text{ and }\\
    \hat{\mathbf{B}}_{l,t,f} &= \sum_{p=1}^P \mathbf{E}_{l,p,t,f} \mathbf{X}_{p,t,f,},
\end{align}
with $\mathbf{E}\in\mathbb{C}^{{(N+1)^2}\times{P}\times{F}\times{T}}$ and $l=1,...,(N+1)^2$. 
The $\text{DNN}$ consists of the following learnable components: a signal encoder, $\text{Enc}_{\text{sig}}$, a directivity encoder, $\text{Enc}_{\text{dir}}$, analysis and synthesis stages (implemented with attention), and a decoder, $\text{Dec}$. Figure~\ref{fig:dnn overview} presents a sketch of our proposed method.

$\text{Enc}_{\text{sig}}$ takes as an input $\mathbf{X}$ and outputs a rich feature representation $\mathbf{Z_X} \in \mathbb{R}^{F \times T \times C}$ of $C$ feature maps for $\mathbf{X}$. To do so, real and imaginary parts of $\mathbf{X}$ are separated to individual feature channels and all subsequent processing is real-valued, i.e., $\mathbf{X}$ is turned to $\mathbf{X}_\mathbb{R}\in\mathbb{R}^{2P \times F \times T}$. $\mathbf{X}_\mathbb{R}$ is then given as input to a series of convolutional blocks, each comprising a 2D convolution, normalization, non-linearity, and dropout. The blocks increase the channel number while preserving other dimensions. After the final block the output dimensionality is $C\times T\times F$ with $C > 2P$. Then, permutation and reshaping operations are used to rearrange the dimensionality according to $\mathbf{Z_X}$.

$\text{Enc}_{\text{dir}}$ takes as an input $\mathbf{H}$ and outputs $\mathbf{Z_H} \in \mathbb{R}^{F \times D \times C}$. Similar to the signal encoder, it first converts complex values to real representation. This is followed by a transposed 2D convolution that increases the channel count and upsamples the frequency dimension $F_H$ to match $F$ from $\mathbf{X}$, ensuring efficient feature learning and dimensional compatibility with the later analysis and synthesis stages. The convolution maintains a receptive field of one in the dimension of $D$. The output is then permuted and reshaped, similarly to the output of $\text{Enc}_{\text{sig}}$, followed by normalization and non-linearity.

Then, a multihead-attention~\cite{vaswani2017attention} produces from $\mathbf{Z_X}$ $F$ query projections $\mathbf{Q}\in\mathbb{R}^{{T}\times{C}}$, and from $\mathbf{Z_H}$ $F$ key ($\mathbf{K}$) and value ($\mathbf{V}$) projections $\in\mathbb{R}^{D\times{C}}$, and outputs ${\mathbf{Z_{Attn}}\in\mathbb{R}^{{F}\times{T}\times{C}}}$. Given the definition of attention~\cite{vaswani2017attention}, 
\begin{equation}
\text{Attention}(\mathbf{Q},\mathbf{K},\mathbf{V}) = \text{softmax}(\mathbf{QK}^{\text{T}}/\sqrt{T})\mathbf{V},
\end{equation}
the scaled dot products of $\mathbf{Q}$ and $\mathbf{K}$ perform an analysis mapping of the directivity features and compares them to the signal features to assess how spatial characteristics within the signal features correspond to the directivity ones.
The product between analysis result (i.e., 
$\text{softmax}(\mathbf{QK}^{\text{T}}/\sqrt{T})$) 
and $\mathbf{V}$ perform a synthesis mapping to the directivity features and creates a linear combination of them to produce a new feature that describes the spatial sound using data derived solely from the array directivities.

$\text{Dec}$ takes as an input $\mathbf{Z_{Attn}}$ and outputs the real-valued matrix
${\mathbf{E_R}\in\mathbb{C}^{{(N+1)^2}\times{2P}\times{F}\times{T}}}$. Similarly to $\mathbf{X}$ and $\mathbf{X_R}$ in $\text{Enc}_{\text{sign}}$, $\mathbf{E_R}$ is mapped to 
the decoding filter matrix $\mathbf{E}$ using reshaping operations.
$\text{Dec}$ is implemented by a learnable affine transform on the C dimension followed by transposing and reshaping to achieve the desired output shape.
\vspace{-4pt}
\section{Evaluation}
\label{sec:evaluation}
\vspace{-4pt}
\subsection{Baseline Methods}
\label{sec:baseline}
\vspace{-4pt}
The proposed method is evaluated against three baseline methods: a static encoding filter matrix, a parametric digital signal processing (DSP) method, and a neural method that receives microphone locations as input metadata.

The first baseline method is based on~\cite{moreau20063d}. It generates a time-invariant encoding filter matrix by minimizing the least-squares error between ideal and measured Ambisonics responses using array directivities. For fair comparison, amplification was not limited.
The second baseline is a parametric DSP method for arbitrary microphone arrays~\cite{mccormack2022parametric}. It involves analysis, i.e. finding  directions of arrival (DOAs), and synthesis of source and ambience ambisonic signals using spatial filters informed by the analyzed DOAs and the device ATFs. For evaluation, only the synthesis part was implemented using oracle DOAs, providing an upper bound on its performance.
The third baseline is Gen-A~\cite{heikkinen2025gen}, a DNN-based method that encodes MA signals to Ambisonics using microphone location metadata. It assumes free-field conditions and omnidirectional microphones, employing a U-Net architecture with separate encoders for signal and metadata. Gen-A was re-trained on the same dataset for a fair comparison.
\vspace{-4pt}
\subsection{Data}
\label{sec:data}
\vspace{-4pt}

Training and evaluation datasets were generated through simulations of multiple sound sources in a room with a recording MA, incorporating both MA ATFs and room acoustics. Two types of ATFs were created: one modeling mobile phone body scattering and another for free-field conditions.

For mobile phone-like arrays, 3D rectangular meshes (four sizes) were used for Boundary Element Method (BEM) simulations. Surface pressure distributions were calculated from 924 incident plane waves at discrete frequencies corresponding to a 128-point  fast Fourier transform (FFT) at 24 kHz. Four-microphone configurations were generated with minimum and maximum pairwise distance of 0.02~m and 0.18~m, and minimum separation along x, y, and z axes of 0.003~m, 0.015~m, and 0.015~m, respectively. 1100 configurations were used for train/val, and 300 for testing.

Free-field MA ATFs followed the process in~\cite{heikkinen2025gen}, sampling 4-mic arrays randomly from a 0.18~m cube with 24 quantized steps and a 0.02~m minimum mic distance.
The ATF of the $p$th microphone was calculated using
\begin{equation}
     h_p(f,\mathbf{u},\mathbf{r}_p)=e^{j2\pi f \mathbf{u}\cdot \mathbf{r}_p /c}
\end{equation}
\noindent
where vector $\mathbf{u}$ denotes the DOA of the plane wave, $\mathbf{r}_p$ is the microphone position, and c is the speed of sound.

Room simulations, using the image source method, featured 1 or 2 point sources and an MA within rooms of 5-10~m dimensions. Minimum distances were enforced: 4 m from walls for MA, 0.3 m for sources, and 2 m between sources and MA. RT60 ranged from 0.3 to 0.5 s.

The dataset comprised 300 training, 100 validation, and 100 testing scenes. Each scene included 300, 100, and 100 randomly sampled ATFs, respectively, along with an ideal Ambisonics reference signal. Separate datasets were created for BEM and free-field directivities. Audio material from the ESC-50 dataset~\cite{piczak2015esc} was convolved with simulated transfer functions (room acoustics + ATFs) to generate MA and reference Ambisonics signals during training.
\vspace{-4pt}
\subsection{Hyperparameters}
\label{sec:hyperparams}
\vspace{-4pt}
Audio sampling rate was 24 kHz. STFT parameters were set to FFT size 256, frame length 128, and hop size 64, reflecting the limited frequency resolution used in the computationally intensive BEM simulations. DNN hyperparameters are selected to balance minimal working conditions and simple encoder/decoder designs, emphasizing cross-attention's effectiveness, rather than for optimal performance.

$\text{Enc}_{\text{sig}}$ has two CNN layers with a square kernel of size 6, and 64 and 256 output channels, respectively.
$\text{Enc}_{\text{dir}}$
has one layer with a 2D transpose convolution of kernel size (1, 4) and stride (1, 2).
The rectified linear unit and the group normalization with eight groups are used throughout the model, and the multi-head attention has four heads. 
Regarding training, a batch size of 16 and time-domain mean squared error loss are used with the Adam optimizer of 2e-4 learning rate. An early stopping protocol was used to end training when validation loss ceased to improve.
\vspace{-4pt}
\subsection{Metrics}
\label{sec:metrics}
\vspace{-4pt}
We evaluate performance using several objective metrics. SI-SDR~\cite{leroux2019sdr} measures overall signal degradation, while magnitude squared coherence and mean magnitude spectrum error assess spatial and spectral quality, respectively~\cite{heikkinen2025gen}. For perceptual quality assessment, we measure inter-channel level difference (ILD) and coherence from binaural signals, generated using a magnitude least-squares decoder~\cite{hold2023magnitude} with the Fabian HRTF database~\cite{brinkmann2017a}. Equivalent rectangular bandwidth (ERB) weighting is applied to the binaural metrics following~\cite{mccormack2022parametric}.
\vspace{-4pt}
\section{Results and Discussion}
\label{sec:results}
\begin{figure}[!t]
   \centering
   \includegraphics[width=1.0\linewidth]{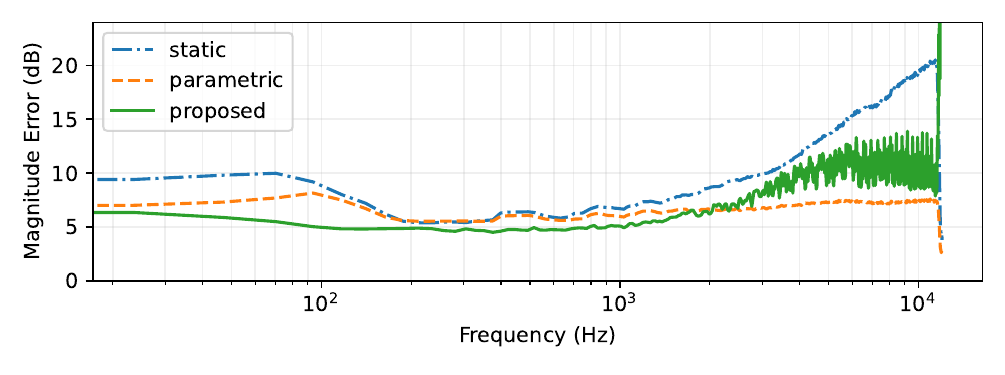}
   \caption{Magnitude response errors on the mobile dataset averaged over Ambisonics channels.}
   \label{fig:magn_response}
\end{figure}

\vspace{-4pt}
Evaluation results are presented in Table~\ref{tab:comparison}. For the mobile phone microphone array (MA) in a complex scattering environment, the proposed method achieved the highest SI-SDR. The parametric method utilizing oracle DOAs demonstrated superior coherence and magnitude spectrum error, as anticipated. This method served as an upper performance anchor due to its inherent model alignment with the simulated scenario of a few direct sources in a reverberant room. The proposed model consistently outperformed the static encoder across all Ambisonics signal metrics, including SI-SDR, coherence, and magnitude spectrum error. Figure \ref{fig:magn_response} shows the average magnitude response errors of the methods highlighting performance of the proposed method under 2 kHz.

In the free-field setting, the proposed method again achieved the best SI-SDR. It also exhibited performance comparable to the neural Gen-A method, specifically designed for this use case. Furthermore, the proposed method surpassed the static encoder in Ambisonics domain signal metrics. The oracle DOA parametric method demonstrated considerably higher performance in coherence and spectral error.
\begin{table}
\centering
\caption{Performance comparison across different methods and datasets}
\vspace{2mm}
\small
\setlength{\tabcolsep}{3pt}
\begin{tabular}{@{}llccccc@{}} 
\multirow{2}{*}{Dataset} & \multirow{2}{*}{Method} & \multicolumn{5}{c}{Metrics} \\
\cline{3-7}
 &  & SI-SDR$\uparrow$ & Coh$\uparrow$ & $\text{MS}_{\text{Err}}$$\downarrow$ & \makecell{Bin\\$\text{ILD}_\text{Err}$}$\downarrow$ & \makecell{Bin\\$\text{Coh}_\text{Err}$}$\downarrow$ \\
\hline
\multirow{3}{*}{\makecell{Mobile\\phone}} & Param. (oracle) & -0.11 & \textbf{0.42} & \textbf{6.95} & \textbf{3.40} & \textbf{0.32} \\
& Static & 0.13 & 0.17 & 13.87 & 4.55 & 0.37 \\ 
 & Proposed & \textbf{1.27} & 0.20 & 9.67 & 4.41 & 0.36 \\
\hline
\multirow{4}{*}{\makecell{Free\\Field}}  & Param. (oracle) & 2.74 & \textbf{0.50} & \textbf{7.44} & \textbf{3.51} & \textbf{0.34} \\
& Static & 4.97 & 0.20 & 14.58 & 4.87 & 0.38 \\
 & Gen-A & 4.66 & 0.20 & 11.32 & 4.63 & 0.38 \\
 & Proposed & \textbf{5.70} & 0.20 & 10.23 & 4.85 & 0.40 \\
\end{tabular}
\label{tab:comparison}
\end{table}

Overall, the proposed method demonstrated strong performance, particularly considering that the depth and number of parameters in its encoders and decoder were intentionally limited. This design choice aimed to highlight the efficacy of cross-attention in addressing the problem. With 890k parameters, the proposed model is less than one-tenth the size of the Gen-A model. Future research could explore performance enhancements by increasing the learning capacity of the encoders and decoder.

The oracle DOA parametric method showed the best performance in binaural metrics for Ambisonics spatial quality. ERB weighting suggests comparable performance among the other methods in frequencies relevant to human hearing, implying that observed differences in Ambisonics metrics may stem from variations in higher frequencies. This requires further investigation and listening tests.

\begin{figure}[!t]
    \centering
    \includegraphics[width=1.0\linewidth]{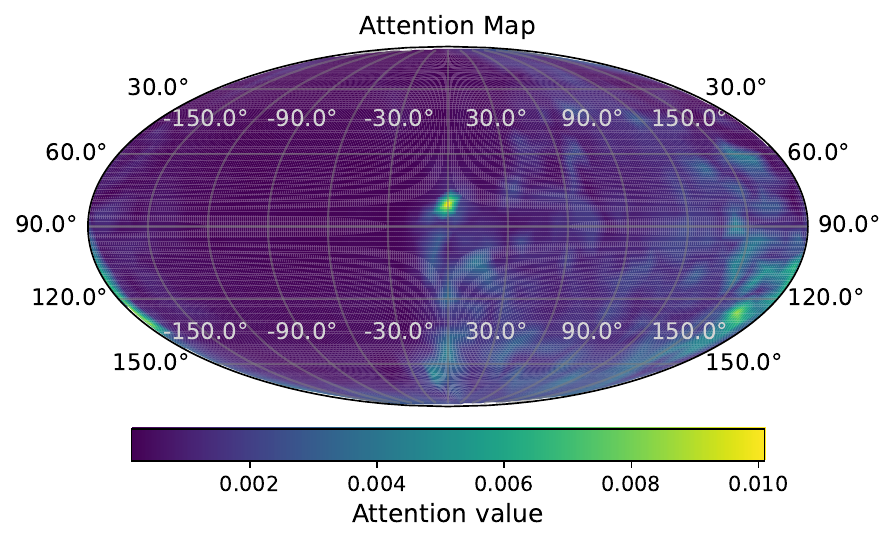}
    \caption{Attention weights averaged over time and frequency with a sound source at 0° azimuth and 90° colatitude in an anechoic room.}
    \label{fig:attn_weights}
\end{figure}

Figure~\ref{fig:attn_weights} illustrates the attention weights for a single sound source positioned directly in front of the MA at a distance of 3.5 m. The largest weight is consistently assigned to a directivity feature closely matching the DOA of the sound. This observation suggests that the model learns to analyze and represent spatial sound using directional information. However, it was also observed that for identical sound source placements with different MAs, the model could assign more broadly distributed attention weights across directional features, without an immediately clear correlation to the source DOA, yet achieving comparable or even superior SI-SDR values. This suggests that the model is capable of learning more complex mappings, warranting further investigation.

\section{Conclusion}
\label{sec:conclusion}
\vspace{-4pt}
We introduced a novel DNN architecture for Ambisonics encoding that utilizes ATFs to enable generalization to unseen geometries and complex scattering environments. A key innovation is the use of cross-attention to combine audio signals with directivity features, creating an array-independent spatial audio representation, with attention weight analysis indicating effective learning of directivity-DOA associations.

Evaluations demonstrated superior performance over baseline solutions, particularly in mobile phone-like scenarios, achieving higher SI-SDR, while performing comparably in free-field conditions. Despite its strong performance, the model maintains a relatively small parameter count, highlighting the efficiency of the cross-attention mechanism. Future work will focus on investigating perceptual quality through additional computational metrics and listening tests, enhancing the learning capacity, and delving deeper into the complex patterns observed in the attention weights.
\vfill\pagebreak
\clearpage
\bibliographystyle{IEEEbib}
\bibliography{strings,refs}
\end{document}